\documentclass[12pt,english,reprint,aps,prb,superscriptaddress]{revtex4-2}
\usepackage[T1]{fontenc}
\usepackage[utf8]{inputenc}
\usepackage{amstext}
\usepackage{amssymb}
\usepackage{graphicx}
\usepackage{adjustbox}
\usepackage{tabularx}
\usepackage{amsmath}
\makeatletter
\usepackage{babel}
\usepackage[unicode=true]{hyperref}
\usepackage{newtxtext}
\usepackage[slantedGreek,varg,varbb]{newtxmath}
\usepackage{mathtools}
\usepackage{siunitx}

\begin{document}

\title {Quantum transport and mobility spectrum of topological carriers in (001) SnTe/PbTe heterojunctions}

\author{D.~Śnieżek}
\affiliation{Institute of Physics, Polish Academy of Sciences, Aleja Lotników 32/46, PL-02-668 Warszawa, Poland}

\author{J.~Wróbel}
\affiliation{Institute of Applied Physics, Military University of Technology, 2 Kaliskiego Str., 00-908 Warsaw, Poland}

\author{M.~Kojdecki}
\affiliation{Institute of Mathematics and Cryptology, Military University of Technology, 2 Kaliskiego Str., 00-908 Warsaw, Poland}

\author{C.~Śliwa}
\affiliation{Institute of Physics, Polish Academy of Sciences, Aleja Lotników 32/46, PL-02-668 Warszawa, Poland}

\author{S.~Schreyeck}
\affiliation{University of Würzburg Am Hubland, Experimental Physics 3, 97074 Würzburg, Germany}

\author{K.~Brunner}
\affiliation{University of Würzburg Am Hubland, Experimental Physics 3, 97074 Würzburg, Germany}

\author{L.~W.~Molenkamp}
\affiliation{University of Würzburg Am Hubland, Experimental Physics 3, 97074 Würzburg, Germany}

\author{G.~Karczewski}
\affiliation{Institute of Physics, Polish Academy of Sciences, Aleja Lotników 32/46, PL-02-668 Warszawa, Poland}

\author{J.~Wróbel}
\email{Corresponding author: wrobel@ifpan.edu.pl}
\affiliation{Institute of Physics, Polish Academy of Sciences, Aleja Lotników 32/46, PL-02-668 Warszawa, Poland}
\affiliation{Institute of Applied Physics, Military University of Technology, 2 Kaliskiego Str., 00-908 Warsaw, Poland}


\begin{abstract}
Measurements of magnetotransport in SnTe/PbTe heterojunctions grown by the MBE technique on (001) undoped CdTe substrates were performed. At low magnetic fields, quantum corrections to conductivity were observed that may be attributed to the presence of topological states at the junction interface. For a sample with 5~nm thick SnTe layer, the data analysis suggests that midgap states are actually gapped. However, the phase coherence effects in 10~nm and 20~nm SnTe/PbTe samples are fully explained assuming existence of gapless Dirac cones. Magnetotransport at higher magnetic fields is described in the framework of mobility spectrum analysis (MSA). We demonstrate that the electron- and hole-like peaks observed simultaneously for all SnTe/PbTe heterojunctions may originate from the concave and convex parts of the energy isosurface for topological states --- and not from the existence of quasiparticles both carrying negative and positive charges. This interpretation is supported by numerical calculations of conductivity tensor components for gapless (100) Dirac cones, performed within a classical model and based on the solutions of Boltzmann transport equation. Our approach shows the feasibility of MSA in application to magnetotransport measurements on topological matter.
\end{abstract}
\maketitle

\section{Introduction}

Topological crystalline insulators (TCIs) are a class of materials in which gapless surface states are protected by crystal mirror symmetry, rather than time-reversal invariance, as in conventional topological insulators (TIs)\cite{Fu2011}. Narrow gap semiconductors SnTe and (Pb,Sn)Te were the first candidates to be declared as members of the new TCI-class. It is known that a band inversion in SnTe occurs at the four $L$ points of the bulk Brillouin zone (BZ), therefore exactly four Dirac cones are expected on the boundary planes (001), (111), and (110) for which the required mirror symmetry \cite{Hsieh2012} is preserved. The gapless surface states, with a linear energy  dispersion, were indeed observed on (001) and (111) surfaces of SnTe-class materials by angle-resolved photo-emission spectroscopy (ARPES) \cite{Dziawa2012, Xu2012}.

In contrast to conventional topological insulators, gapless surface states of TCIs have a much more tunable properties, as it is relatively easy to lower spatial symmetry of a system. The application of electric field \cite{Liu2014} or uni-axial and bi-axial strains can break the symmetry protection and open the gap for Dirac states \cite{Okada2013}. Moreover, the presence of multiple Dirac cones may allow construction of novel quantum devices based on the so-called valleytronics \cite{Zhao2015}. However, the transport studies of topological surface states (TSS) is difficult in SnTe-class materials because of strong $p$-type conductivity of intrinsic holes, with $p \sim 10^{20}\ \mathrm{cm}^{-3}$ \cite{Dybko2017}. Such unintentional doping is provided by electrically active Sn-vacancies \cite{Zemel1965}. As a result, the chemical potential is anchored deep the valence band and conduction through TSS is masked by bulk conductance.

Therefore, to enhance the surface-to-bulk ratio and to move the Fermi energy towards Dirac points, epitaxial growth of thin ($\lesssim 100\ \mathrm{nm})$ SnTe films is usually performed on several insulating and conducting substrates. In particular, the (111) layers were grown on BaF$_2$ \cite{Akiyama2014}, CdTe \cite{Ishikawa2016} and Bi$_2$Te$_3$ \cite{Taskin2014} suporting templates. SnTe films of the (001) crystallographic orientation were deposited on BaF$_2$ \cite{Assaf2014} and SrTiO$_3$ (STO) insulators \cite{Zou2019, Albright2021}. Recenty, the fabrication of CdTe/SnTe/CdTe quantum wells on GaAs substrates \cite{Dybko2018} and SnTe/PbTe heterojunctions on STO templates \cite{Wei2018, Wei2019} were reported. In both cases, the epitaxial layers were grown along the [001] crystallographic direction.

\begin{figure*}
	\begin{centering} 
		\includegraphics{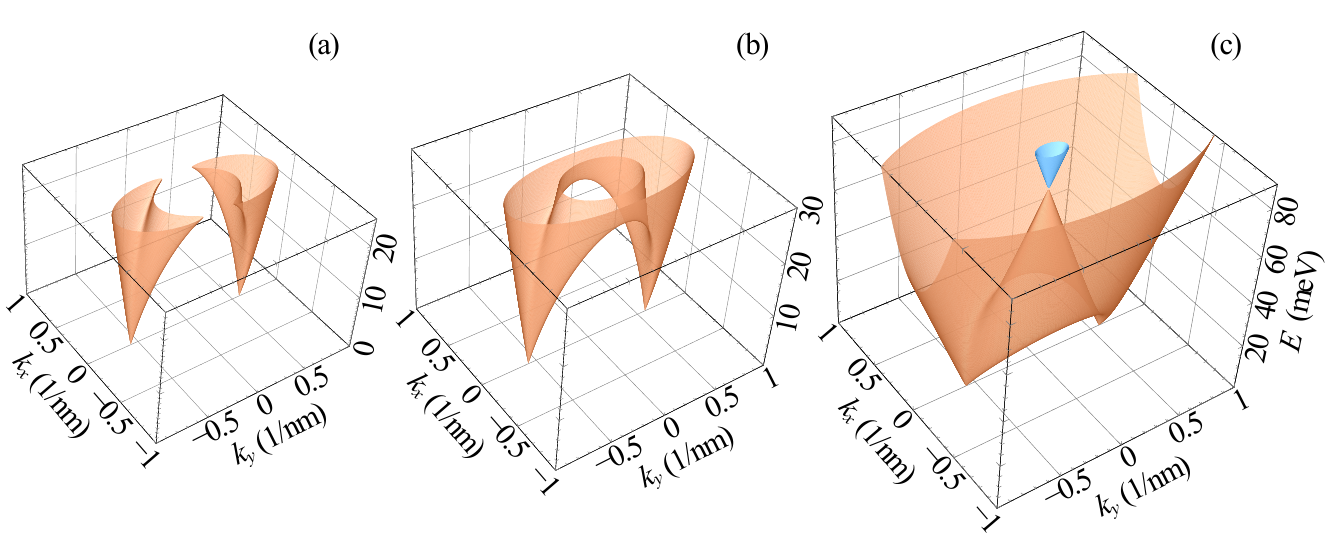} 
	\par\end{centering} 
	\caption{Energy band structure $E_{\mathrm{H,L}}(\mathbf{k})$ for the (001) surface states, $(k_x,k_y)=(0,0)$ corresponds to $\bar{X}$ point of 2D Brillouin zone \cite{Liu2013}. Bands $E_\mathrm{L}$ with positive energies are plotted up to Fermi level $E_\mathrm{F} = 25\ \mathrm{meV}$ (a), $E_\mathrm{F}=30\ \mathrm{meV}$ (b) and $E_\mathrm{F}=90\ \mathrm{meV}$ (c). Subfigures (a) and (b) illustrate the constant energy planes, just below and above a Lifshitz transition, which occurs at $E_\mathrm{F} = 26\ \mathrm{meV}$. For energies $E_\mathrm{F} \gtrapprox 75\ \mathrm{meV}$ the upper band $E_\mathrm{H}$, shown as blue cone in (c), becomes occupied and coexists with bulk states.\label{fig:snte-dirac-cones}}	
\end{figure*}

TSS on the (001) boundary of SnTe are, in the following sense, more interesting than (111) surface states. In the former case, the Dirac points are not located at the $\bar{X}$ point in Brillouin zone, but are slightly shifted along $\bar{\Gamma}\bar{X}$ direction \cite{Hsieh2012}. Moreover, Fermi surface topology changes as a function of energy (Lifshitz transition), and a Van Hove singularity in the density of states (DOS) is present at the energy corresponding to the cones' tip. To discuss the topology of the SnTe (001) states, Liu \textit{et al}.\ \cite{Liu2013} introduced a low-energy $\mathbf{k}\cdot \mathbf{p}$ model near the $\bar{X}$ point of BZ. Some results of model calculations are shown in Fig.~\ref{fig:snte-dirac-cones}, for three different values of Fermi energy $E_\mathrm{F}$.

In this work we report on magnetotransport measurements of SnTe/PbTe heterojunctions, which were grown by molecular beam epitaxy (MBE) on CdTe/GaAs substrates along the [001] crystallographic direction. At low temperatures and low magnetic fields we have observed characteristic corrections to the conductivity, related to the interference of electronic wave functions. Data were analyzed using a modified Hikami, Larkin, and Nagaoka (HLN) model, which describes the quantum coherence effects and contains an additional quadratic term accounting for the classical magnetoresistance \cite{Assaf2013}. 

Classical magnetotransport at higher fields was described using the so-called mobility spectrum analysis (MSA), which is extremely useful in the case of multi-carrier transport \cite{Beck1987}. The MSA method was already used for topological materials \cite{Grabecki2020, Wang2021, Wadge2022}, however,  separate peaks detected in mobility spectra are traditionally interpreted as the presence of distinct transport channels. Here we show, that electron-like and hole-like peaks, observed simultaneously for SnTe/PbTe heterojunctions, originate from the concave and convex parts of constant energy surface of topological states, see Fig.~\ref{fig:snte-dirac-cones}. In other words, both peaks account for the \textit{single-carrier} transport in the \textit{single-band} of TSS states. 

This claim was supported by the theoretical calculations of conductivity tensor for carriers described by a Liu \textit{et al}. model. Tensor components $\sigma_{xx}$ and $\sigma_{xy}$ were obtained numerically by adopting a McClure approach \cite{McClure1956} which is based on the solutions of Boltzmann transport equation. Calculations showed the extremely rich mobility spectra of topological carriers, with a pattern strongly changing with $E_\mathrm{F}$. Therefore, MSA method can in principle be used not only for identification of (001) surface states, but also for localization of Fermi level, relative to Dirac point and Van Hove singularity.
		
\section{Samples preparation}

\begin{figure}
	\begin{centering}
		\includegraphics{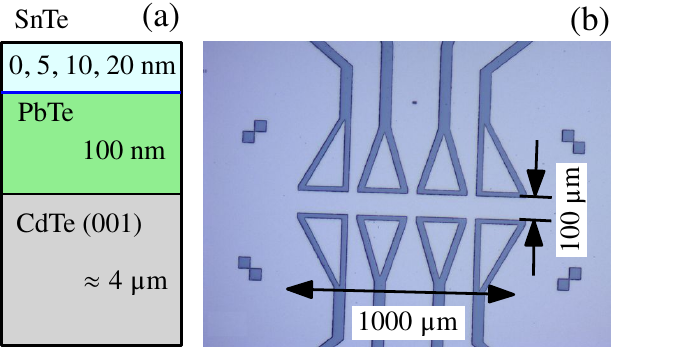}
	\par \end{centering}
	\caption{(a) SnTe/PbTe heterojunction scheme with thicknesses of epilayers indicated. (b) 8-terminal Hall structure patterned using electron beam lithography with conducting channel of $1000\ \muup\mathrm{m}$ length and $100\ \muup\mathrm{m}$ width}
	\label{fig:heterostructure}
\end{figure}

SnTe/PbTe heterostructures were grown by molecular beam epitaxy (MBE) on (001) oriented CdTe undoped substrates. The growth started from covering the substrate by a few micrometer of epitaxial CdTe, then by depositing a $100\ \mathrm{nm}$ thick PbTe layer and subsequently, a SnTe film of varying thickness ($0\ \mathrm{nm}$, $5\ \mathrm{nm}$, $10\ \mathrm{nm}$ and $20\ \mathrm{nm}$), see Fig.~\ref{fig:heterostructure}a. In order to keep the two layers homogeneous and prevent mixing of Pb and Sn, MBE growth was carried out at the lowest possible substrate temperature of $230\ ^{\circ}\mathrm{C}$, and the sample was cooled immediately after growth. The entire process was controlled by high-energy electron diffraction (RHEED) showing the excellent quality of the successive layers and that they preserve the orientation of (001) substrate \cite{*[{}] [{ and references therein.}] Chusnutdinow2020}. Based on the oscillations of the RHEED signal, we accurately determined and controlled the thicknesses of PbTe and SnTe layers. Moreover, during the growth of PbTe, the Pb/Te flux ratio was adjusted to assure \textit{n--}type conductivity with an electron concentration of the order of $10^{18}\ \mathrm{cm}^{-3}$ . The conductivity of SnTe is always \textit{p--}type, as already noted and this way we are able to fabricate a \textit{p--n} heterojunctions.

Samples of sizes $5\times 5$~mm were cut from such wafers and covered with e-beam resist PMMA. For further processing we have applied the low-temperature method, developed earlier for II-VI semiconductors quantum wells \cite{Majewicz2014}. In particular, samples were baked for $1$ hour in $120\ ^{\circ}\mathrm{C}$ at lowered pressure to avoid material damage, reduce interdiffusion and assure a better drying of the resist film. Using electron beam lithography we have patterned 8-terminal Hall bar devices with conducting channel size of $1$~mm length and $100\ \muup\mathrm{m}$ width (Fig.~\ref{fig:heterostructure}b). After developing, pattern was etched in $0.06 \%$ Br$_{2}$ solution in ethylene glycol for $5$ minutes achieving $\approx 330$~nm depth of mesas. Macroscopic contacts were made with silver paint and connected in parallel to both SnTe and PbTe layers. From the same wafers we prepared 4-terminal square samples in van der Pauw geometry, not using lithography and avoding thermal post-processing. Test measurements showed that the low-temperature-method applied for fabrication of Hall-bar samples did not deteriorate the electrical properties of SnTe/PbTe epilayers.	
				
\section{Magnetotransport measurements} \label{results}

\begin{figure*}
	\includegraphics[width=\linewidth]{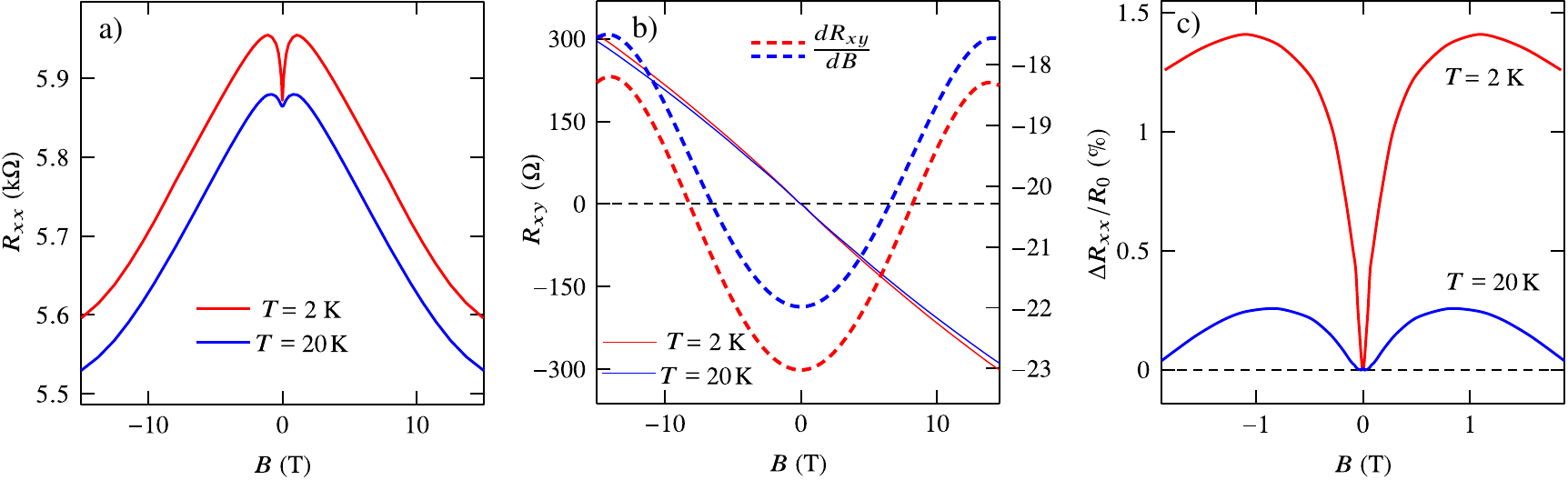}
	\caption{Magnetotransport measurements for PbTe layer ($0$~nm SnTe) at  $T=2$~K and $20$~K  (a) Longitudinal magnetoresistance $R_{xx}$ vs magnetic field $B$. (b) Hall resistance $R_{xy}$ (left axis) together with derivatives $\mathrm{d}R_{xy}/\mathrm{d}B$ (right axis). (c) Relative changes of resistance $\Delta R_{xx}/R_{0}$ at low magnetic fields, where $R_0=R_{xx}(0)$}
	\label{fig:pbte_measurements}
\end{figure*}

\begin{figure*}
	\begin{centering}
		\includegraphics[width=\linewidth]{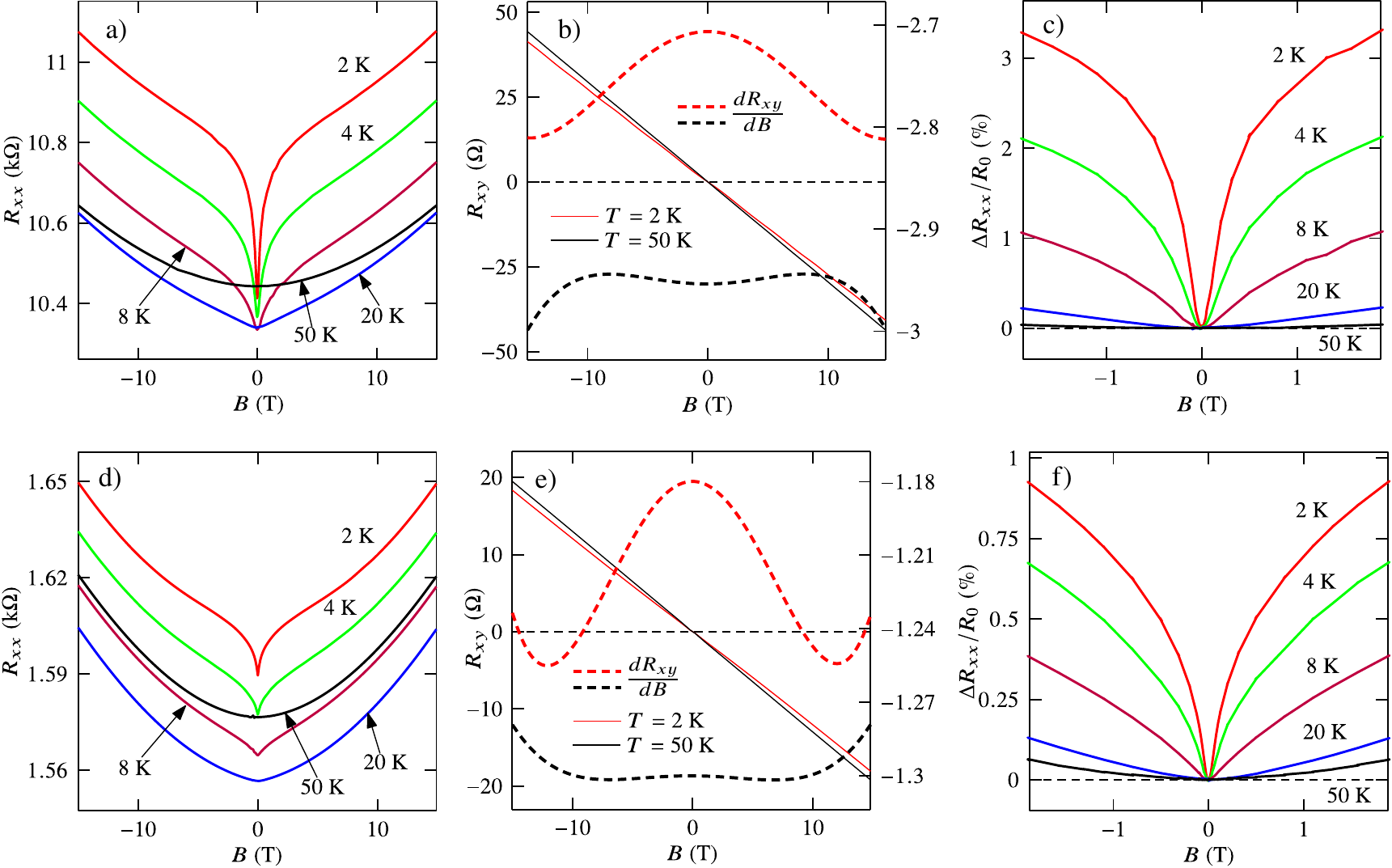}
	\par \end{centering}
	\caption{Results for $5$~nm (a,b,c) and $20$~nm (d,e,f) SnTe layers at $T=2\ \mathrm{K},\ 4\ \mathrm{K},\ 8\ \mathrm{K},\ 20\ \mathrm{K}\ \mathrm{and}\ 50\ \mathrm{K}$. (a,d) Longitudinal magnetoresistance $R_{xx}$ vs magnetic field $B$. (b,e) Hall resistance $R_{xy}$ and derivatives $\mathrm{d}R_{xy}/\mathrm{d}B$ (dashed lines) at $T=2\ \mathrm{K}\ \mathrm{and}\ 20\ \mathrm{K}$. (c,f) Relative changes of resistance $\Delta R_{xx}/R_{0}$ at low magnetic fields.}
	\label{fig:snte_5_20_measurements}
\end{figure*}

\begin{figure*}
	\includegraphics[width=\linewidth]{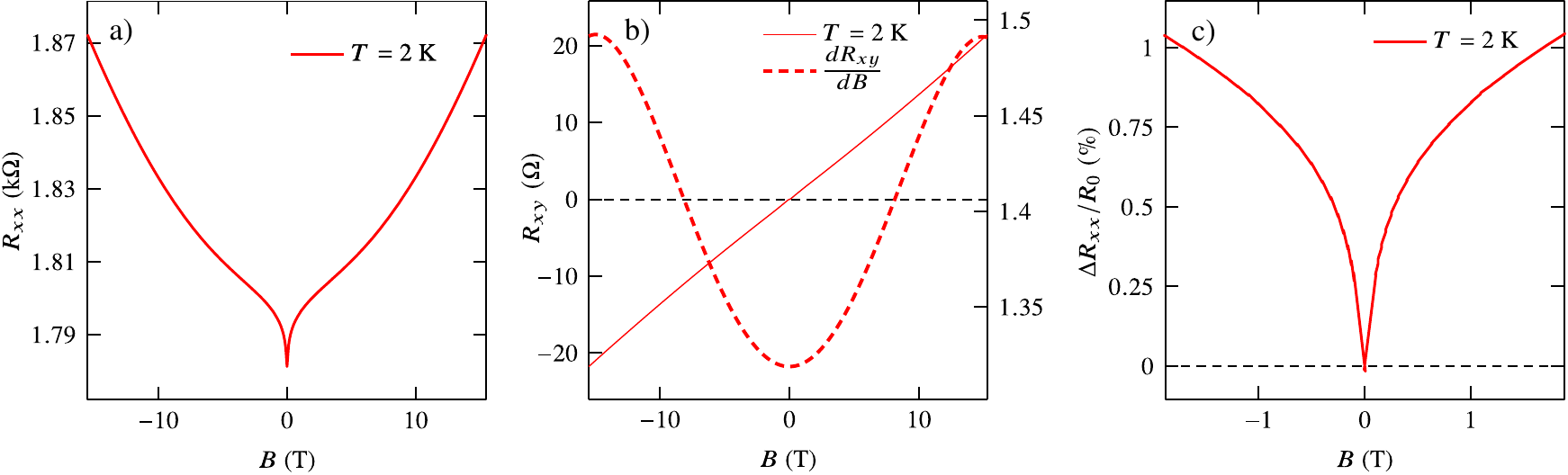}
	\caption{Results for $10$~nm SnTe layer at $T=2$~K. (a) Longitudinal magnetoresistance $R_{xx}$  vs B. (b)  Hall resistance $R_{xy}$ and derivatives $\mathrm{d}R_{xy}/\mathrm{d}B$ (dashed lines). (c) Relative changes of resistance at low magnetic fields.}
	\label{fig:snte_10_measurements}
\end{figure*}

Patterned Hall bar devices were measured in He4 cryostat at magnetic fields $B$ up to $B_{\mathrm{max}}=15$~T, using the constant current (DC) mode. For 10~nm and 20~nm SnTe/PbTe layers, we used the excitation current of $50\ \muup\mathrm{A}$. For 5~nm SnTe/PbTe junction and for single PbTe layer, which was not covered with SnTe, the smaller current of $5\ \muup\mathrm{A}$ was applied. We measured the longitudinal $R_{xx}(B)$ and vertical (Hall) $R_{xy}(B)$ resistances at temperatures $T=2\ \mathrm{K},\ 4\ \mathrm{K},\ 8\ \mathrm{K},\ 20\ \mathrm{K}\ \mathrm{and}\ 50\ \mathrm{K}$ for 5~nm and 20~nm SnTe/PbTe junctions. For PbTe layer and 10~nm SnTe/PbTe sample measurements were performed at $T=2\ \mathrm{K}\ \mathrm{and}\ 20\ \mathrm{K}$ only. Data were collected for both directions of magnetic field and symmetrized at $\pm B$ points, in order to remove contact asymmetry effects. Results are shown in Figs.~\ref{fig:pbte_measurements}, \ref{fig:snte_5_20_measurements} and \ref{fig:snte_10_measurements}.

The Hall resistance data that we show, include also the slopes of smoothed $R_{xy}(B)$ curves, which change with magnetic field, indicating the presence charge carriers with different mobilities. For the mobility spectrum analysis (MSA) of multi-carrier transport in 2D, we have calculated conductivity tensor components using standard formulas $\sigma_{xx} = R_{xx}/(R_{xx}^2 + R_{xy}^2)$ and $\sigma_{xy} = R_{xy}/(R_{xx}^2 + R_{xy}^2)$. Additionally, on the right hand side of the figures, we indicated the relative changes of longitudinal resistance $R_{xy}(B)$ at low magnetic fields. Clearly, the narrow minima, caused by weak anti-localization WAL) are visible for all samples. For PbTe layer, not only WAL, but also the characteristic cusp, induced by weak localization (WL) was observed. It seems that in the case of PbTe layer, which was not covered by SnTe, quantum corrections to conductivity dominated the whole range of magnetic fields, see Fig.~\ref{fig:pbte_measurements}a. 
		
\section{Quantum corrections to conductivity}

The quantum corrections to the magnetoconductace in 2D systems with strong spin-orbit coupling are commonly described by Hikami-Larkin-Nagaoka (HLN) model \cite{Hikami1980}, which can be written as follows \cite{Assaf2013}:

\begin{equation}
	\Delta G(B)=\eta\,\Delta G_{1}+\eta\,\Delta G_{2}-\beta B^{2},\label{eq:HLN}
\end{equation}
where
\begin{equation}
	\Delta G_{1}=\frac{\alpha e^{2}}{\pi h}\left[\psi\left(\frac{B_{\phi}}{B}+\frac{1}{2}\right)-\ln\left(\frac{B_{\phi}}{B}\right)\right],\label{eq:DGphi}
\end{equation}
\begin{equation}
	\Delta G_{2}=-\frac{3\alpha e^{2}}{\pi h}\left[\psi\left(\frac{4B_{\mathrm{SO}}+3B_{\phi}}{3B}+\frac{1}{2}\right)
	-\ln\left(\frac{4B_{\mathrm{SO}}+3B_{\phi}}{3B}\right)\right].\label{eq:DGSO}
\end{equation}

Here $\psi$ is the digamma function and $\alpha=-1/2$ for the so-called symplectic class. Therefore, the first term is responsible for weak anti-localization (WAL), the second one for weak localization (WL) effects. The formula involves characteristic fields $B_{\phi}=\hslash/(4eL_{\phi}^{2})$ and $B_{\mathrm{SO}}=\hslash/(4eL_{\mathrm{SO}}^{2})$ for scattering channels, $L_{\phi}$ and $L_{\mathrm{SO}}$ standing for the phase coherence and spin-orbit lenghts, respectively. Both terms are multiplied by the parameter $\eta>0$, which accounts for the \emph{effective number of quantum channels} contributing to transport.

In general, HLN model contains also the term with $B_{\mathrm{e}}=\hslash/(4e\ell_{\mathrm{e}}^{2})$, where $\ell_{\mathrm{e}}$ is the elastic scattering length. We estimated, however, that for electrons in PbTe and holes in SnTe the mean free paths $\ell_{e}$ are very short and fall within the range $1.0\ \mathrm{nm}$ to $3.0\ \mathrm{nm}$, see Sec.~\ref{sec:Mobility-spectrum-of-junctions}. Therefore, for all samples terms containing $B_{\mathrm{e}}\gg B_{\mathrm{max}}=15\ \mathrm{T}$ are negligible and can be safely ignored. Instead, following \cite{Assaf2013}, we have modified HLN formula by including an additional quadratic term $\beta B^{2}$, which accounts for the \emph{classical positive magnetoresistance}, of the type described by formula \ref{eq:rhoxx}, given below. We did not expect, however, that such a single term describes correctly the classical magnetoresistance in the entire range of fields, because of multi-carrier transport and non-spherical shape of Fermi surfaces.

Therefore, we have used formula (\ref{eq:HLN}) to fit $\Delta G(B)$ to the experimental data for magnetic fields $B<3\ \mathrm{T}$ only. Characteristic lengths $L_{\phi}$ and $L_{\mathrm{SO}}$, together with coefficients $\eta$ and $\beta$, were treated as the fitting parameters. As it has been already noticed \cite{Peres2014}, the complex nature of HLN model often leads to results which depend on the initial guess, if the standard Levenberg-Marquard optimization techniques are used. In our case, we applied the so-called \emph{covariance matrix adaptation evolution strategy} (CMA-ES) --- an iterative method, where in each iteration a new candidate solutions are generated in a stochastic way \cite{Hansen2016}. We hope, that this kind of numerical optimization procedure makes the solutions less dependent on the initial values of the parameters being fitted. We used for calculations the implementation of the CMA-ES algorithm which is available in the Python language environment \cite{Nomura2020}.

\subsection{PbTe layer}

\begin{figure}
	\begin{centering}
		\includegraphics{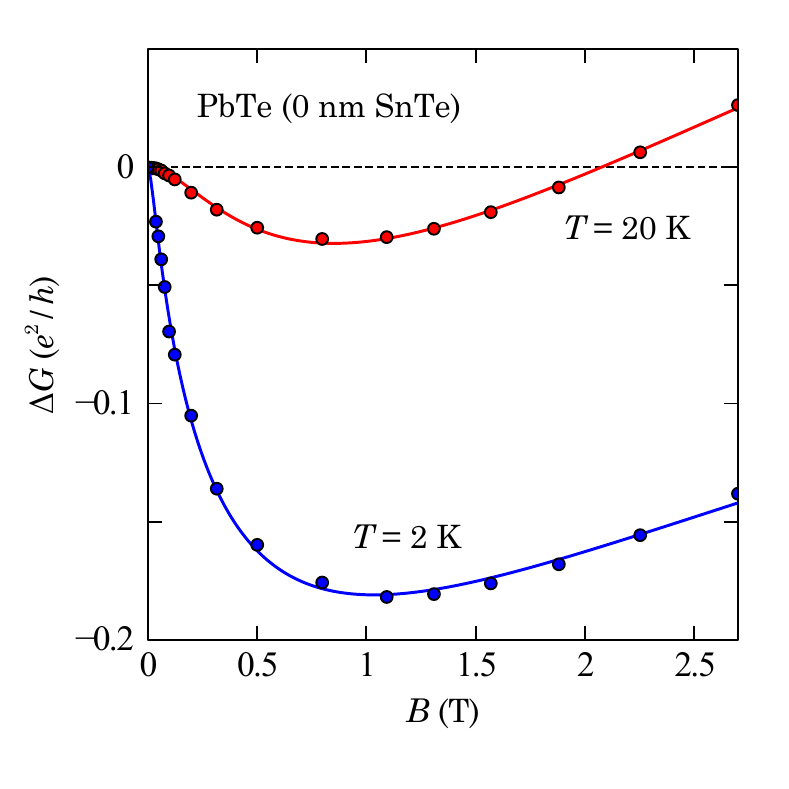}
		\par\end{centering}
	\caption{Measured conductance corrections (points) and fitted curves (lines)
		for PbTe layer at temperatures $2\ \mathrm{K}$ and $20\ K$. The
		following parameters were obtained: $\eta=0.61$, $L_{\phi}=120.0\ \mathrm{nm}$,
		$L_{\mathrm{SO}}=27.3\ \mathrm{nm}$, $\beta=\num{1.0e-6}$ for $T=2\ \mathrm{K}$
		and $\eta=0.97$, $L_{\phi}=37.2\ \mathrm{nm}$, $L_{\mathrm{SO}}=27.0\ \mathrm{nm}$,
		$\beta=\num{1.3e-5}$ for $T=20\ \mathrm{K}$ (here $\beta$ is in
		$e^{2}/h$ per tesla squared units).\label{fig:Fit PbTe}}
\end{figure}

Results of the fitting procedures, applied to PbTe layer data, are shown in Fig.~\ref{fig:Fit PbTe}. We note, that values of the parameter $\beta$ are rather small, which confirms that the classical contribution becomes important only at higher fields and that the $\Delta G(B)$ data for $B>3\ \mathrm{T}$ are well described by quantum terms only. Data show that phase coherence length $L_{\phi}$ decreased with temperature approximately as $T^{-0.5}$. This may suggest the electron-electron collisions as a phase decay mechanism\cite{Altshuler1982}, however, two data points are probably not enough to definitely identify the dominating inelastic process.

Inelastic processes are important also for the determination of effective number of quantum channels contributing to transport. For PbTe layers grown along the {[}001{]} direction we expect $\eta=4$, which is the number of equivalent constant energy ellipsoids at $L$ points of Brillouin zone. In our case the value of the parameter $\eta$ is smaller, however, it increases with temperature from $0.61$ at $2\ \mathrm{K}$ to $0.97$ at 20K. At the same time, phase coherence length decreases from 120.0~nm to 37.2~nm, which strongly suggests that the reduction of the parameter $\eta$ is caused by some coherent processes, which is less effective at higher temperatures.

Most probably, such renormalization results from electron scattering between equivalent elipsoids \cite{Fukuyama1980}. For the first time, the intrasurface valley coupling, which was responsible for reducing parameter $\eta$ from 2 to approximately 1, was observed for Si inversion layers, which are the two-valley systems\cite{Kuntsevich2007}. In our case, however, at low temperature $\eta<1$, which indicates some additional contribution to quantum corrections, which most probably originates from the weak localization of bulk electrons \cite{Lu2011}. Indeed, at $T=2\ \mathrm{K}$ the phase coherence length $L_{\phi}$ is larger than the total thickness of PbTe layer ($100\ \mathrm{nm}$) Therefore, we expect some additional quantum interference effects, characteristic for mesoscopic systems. At $T=20\ \mathrm{K}$ $L_{\phi}$ is reduced and $\eta$ approaches $1$.

As opposed to $L_{\phi}$, the spin-orbit length $L_{\mathrm{SO}}$ does not change much with temperature. This result suggests, that the Rashba effect is responsible for spin-orbit interactions in 2D system confined on PbTe surface. According to \cite{Bychkov1984}, the zero field spin-splitting energy is given by $\Delta_{\mathrm{SO}}=\alpha_{\mathrm{R}}k_{\mathrm{F}}$, where $\alpha_{\mathrm{R}}$ is the coupling constant and $k_{\mathrm{F}}$ is the quasi-momentum at the Fermi surface. On the other hand, $L_{\mathrm{SO}}\propto\Delta_{\mathrm{SO}}^{-1}$, therefore the spin-orbit length does not change, if the carrier density is constant as a function of temperature, since $k_{F}=\sqrt{2\pi n}$. This is indeed the case for $2\ \mathrm{K}$ to $20\ \mathrm{K}$ range, as it is confirmed by mobility spectrum analysis, see Sec. \ref{sec:Mobility-spectrum-of-junctions}. The independent on temperature spin-orbit length $L_{\mathrm{SO}}$ has been already reported for PbTe quantum wells grown on (111) planes \cite{Peres2014}.

\subsection{SnTe/PbTe junctions}

\begin{figure*}
	\begin{centering}
		\includegraphics{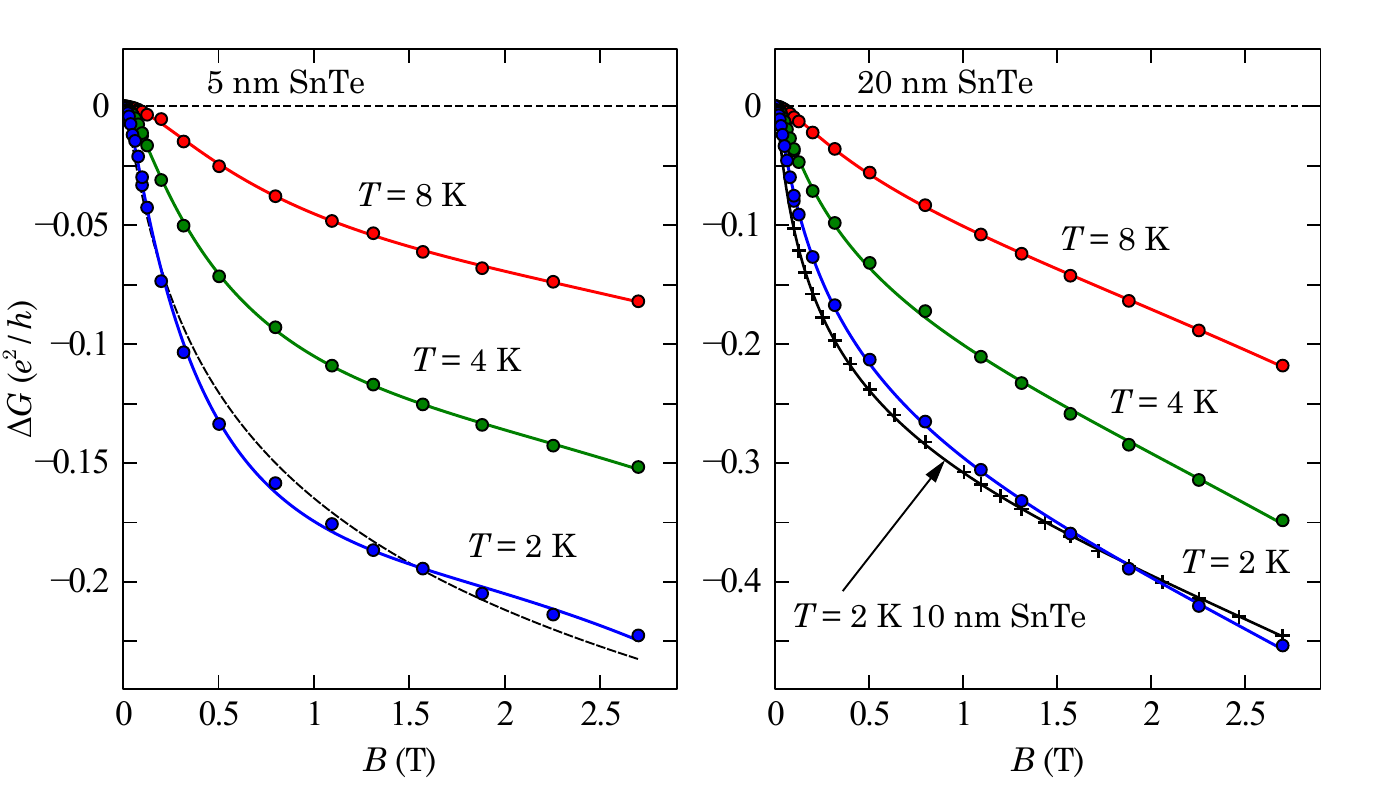}
		\par\end{centering}
	\caption{Conductance corrections $\Delta G(B)$ for 5~nm SnTe junction (left)
		and 20~nm SnTe (right) at low temperatures. Data for 10 nm SnTe junctio (at $T=2\ \mathrm{K}$) are also included. Measured values are depicted	with symbols, fitted curves with lines. Obtained parameters are summarized in table~\ref{tab:fitpar}.
		 \label{fig:fit_SnTe}}
\end{figure*}

The weak anti-localization effect, commonly observed for topological surface states (TSS), arises from the accumulation of Berry phase $\pi$ by helical carriers with spin-momentum locking. Therefore it is assumed, that quantum corrections to the conductance of gapless Dirac fermions are simply given by formula (\ref{eq:DGphi}), with the same value of $\alpha=-1/2$ \cite{Garate2012}. For a single SnTe surface, the number of Dirac cones $\eta=4$ for both $(100)$ and $(111)$ orientations. The results of fitting conductance corrections $\Delta G(B)$ for SnTe/PbTe junction with HLN model are shown in Fig. \ref{fig:fit_SnTe}.

\begin{figure}
	\begin{centering}
		\includegraphics{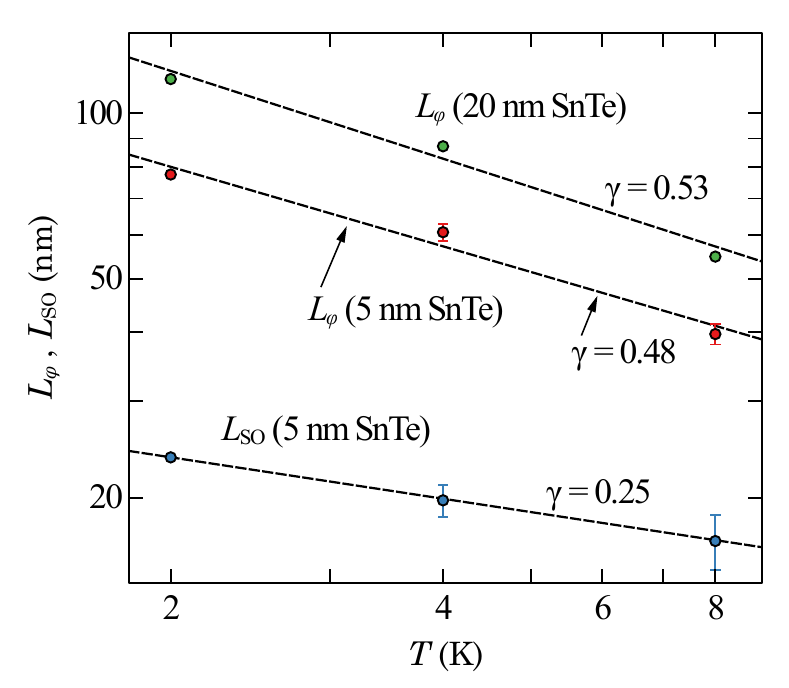}
	\par\end{centering}
	\caption{Characteristic lengths obtained as a fitting parameters to HLN model for 5\ nm SnTe/PbTe and 20\ nm SnTe/PbTe samples as a function vs temperature $T$. Dashed lines represent estimated $T^{-\gamma}$ decay. \label{fig:LPhi_LSO_vs_T}}
\end{figure}

Initially, we fitted the magnetoconductance data to the simplest version of a modified HLN formula
\begin{equation}
	\Delta G(B)=\eta\,\Delta G_{1}-\beta B^{2}\label{eq:HLN-Dirac},
\end{equation}
where $\Delta G_{1}$ given by Eq.~\ref{eq:DGphi}, which is commonly used for TSS. For 5\ nm SnTe/PbTe junction, however, the fit quality was not satisfactory as it is shown with the dashed line in Fig.\ \ref{fig:fit_SnTe}. Therefore, data obtained for 5\ nm SnTe sample have been fitted to Eq.\ \ref{eq:HLN}. In other words, we have used the same procedure as for PbTe sample. The fits are shown with solid lines and fitted parameters are shown in table~\ref{tab:fitpar} and plotted as a function of temperature in Fig.~\ref{fig:LPhi_LSO_vs_T}. 
\begin{table}[h!]
	\caption{\label{tab:fitpar}}
	\begin{ruledtabular}
		\begin{tabular}{lccccc}
			SnTe & $T$ (K) & $\eta$ & $L{\phi}$ (nm) & $L_{\mathrm{SO}}$ (nm) & $\beta$ (T$^{-2}$) \\
			\colrule
			$5$ nm &	$2$ & $0.78$ & $77.4$ & $23.7$ & \num{6.6e-4}\\
			&	$4$ & $0.57$ & $60.8$ & $19.8$ & \num{4.8e-4}\\
			&	$8$ & $0.45$ & $39.7$ & $16.7$ & \num{3.5e-4}\\
			& & & & & \\
			$10$ nm &	$2$ & $0.61$ & $159.8$ & $-$ & \num{5.1e-4}\\
			& & & & & \\
			$20$ nm &	$2$ & $0.73$ & $115.4$ & $-$ & \num{5.6e-4}\\
			&	$4$ & $0.61$ & $87.1$ & $-$ & \num{5.6e-4}\\
			&	$8$ & $0.49$ & $54.9$ & $-$ & \num{5.7e-4}\\
		\end{tabular} 
	\end{ruledtabular}
\end{table}
Similarly to 2DEG on PbTe surface, the phase coherence length $L_{\phi}$ decreased with temperature approximately as $T^{-0.5}$. Contrary to PbTe sample, however, parameter $L_{\mathrm{SO}}$ also decreased with $T$, but the decay was slower in comparison to $L_{\phi}$. At low temperatures $L_{\mathrm{SO}}$, obtained for 5\ nm sample, decreased approximately as $T^{-0.25}$, see Fig.\ \ref{fig:LPhi_LSO_vs_T}.

For 10\ nm and 20\ nm SnTe/PbTe junctions, the parameter $L_{\mathrm{SO}}$ is not shown in figures, since for those samples the formula (\ref{eq:HLN-Dirac}) was fully sufficient. Adding $\Delta G_{2}$ term did not change the fit quality, therefore the use of Eq.\ \ref{eq:HLN} was not justified. This shows, that from the point of view of quantum corrections, the 5\ nm sample is in some intermediate position between a single PbTe layer and 10\ nm and 20\ nm SnTe/PbTe junctions. Nevertheless, the effective number of quantum channels, $\eta$ is less then $1$ for all junctions and decreases with temperature in the studied range. For example, in 20 nm SnTe sample, $\eta=0.73$ at $T=2\ \mathrm{K}$ and $\eta=0.49$ at $T=8\ \mathrm{K}$. As already discussed, the reduction of number of channels below expected limit $\eta=1$ is most probably caused by the WL contribution coming from bulk carriers.

In order to verify if this hypothesis holds for 10\ nm and 20\ nm SnTe samples, we performed an alternative fit of $\Delta G(B)$ data to formula $\Delta G(B)=\eta_{1}\Delta G_{1}^{\mathrm{WAL}}+\eta_{2}\Delta G_{1}^{\mathrm{WL}}$, which adds the $\Delta G_{1}^{\mathrm{WAL}}$ contribution from topological carriers and the $\Delta G_{1}^{\mathrm{WL}}$ part from trivial bulk states\cite{Akiyama2014, Albright2021}. Here $\Delta G_{1}$ is again given by Eq.\ \ref{eq:DGphi}, $\alpha=-1/2$ for WAL as before and $\alpha=+1$ for WL contibution. The effective numbers of channels $\eta_{1}$, $\eta_{2}$ and phase coherence lengths $L_{\phi}^{\mathrm{WAL}}$, $L_{\phi}^{\mathrm{WL}}$ were treated as fitting parameters. It turned out, however, that final results were very sensitive to the initial values of parameters used by the fit procedure. Nevertheless, from multiple runs of CMA-ES algorithm we were able to estimate that $\eta_{1}\approx1$. Therefore, for the alternative fit we used
\begin{equation}
	\Delta G(B)=\Delta G_{1}^{\mathrm{WAL}}+\eta_{2}\Delta G_{1}^{\mathrm{WL}}-\beta B^{2},\label{eq:Alt-fit}
\end{equation}
i.e. we fixed $\eta_{1}=1$ for topological states. Final results were practically undistinguishable from the earlier fits to formula \ref{eq:HLN-Dirac}, which are shown in Fig.\ \ref{fig:fit_SnTe}. For example, we obtained $L_{\phi}^{\mathrm{WAL}}=141.8\ \mathrm{nm}$, $\mathrm{\eta_{2}=0.20}$ and $L_{\phi}^{\mathrm{WL}}=115\ \mathrm{nm}$ from magnetoconductivity data of 10\ nm SnTe sample at $T=2\ \mathrm{K}$. As expected, at low temperatures $L_{\phi}^{\mathrm{WL}}$ is larger than the thickness of SnTe epilayer.

We conclude that quantum corrections to conductance for 10\ nm and 20\ nm SnTe samples are well described by HLN model for gapless fermions with the effective number of channels $\eta<1$ or, alternatively, by the same model with $\eta=1$ and the additional WL contribution from bulk carriers. The latter approach delegates the $\eta<1$ problem from topological carriers to trivial states in the bulk. Indeed, from the alternative fit to 20\ nm SnTe sample data we obtained $\eta_{2}=0.14$, $0.20$ and $0.29$ at temperatures $T=2\ \mathrm{K}$, $4\ \mathrm{K}$ and $8\ \mathrm{K}$ respectively. The effective number of quantum channels, which is less than $1$ for the trivial states, was already reported for thin SnTe films and is attributed to the band edge fluctuations on the epilayer plane \cite{Albright2021}.

Finally, we mention that the $\beta$ parameter depends rather weakly on the temperature for all SnTe/PbTe junctions and falls within the $\num{3.5e-4}$ to $\num{6.6e-4}$ range for $T<8\ \mathrm{K}$. This demonstrates that the classical contribution $\beta B^{2}$ is much larger for SnTe/PbTe samples then for single PbTe layer, suggesting the presence of the multi-carrier transport --- which is actually expected for \emph{p--n} junctions and which is revealed by the analysis of the classical magnetoresistance data.

\section{Mobility spectrum}

Classical transport in topological materials is usually analyzed using Drude expressions for $\sigma_{xx}(B)$ and $\sigma_{xy}(B)$ conductivity tensor components of each charge carrier type \cite{Assaf2014,Akiyama2014,Ishikawa2016,Wei2019}. However, this approach is not valid for lead chalcogenides, whose bands are characterized by highly anisotropic effective masses. In particular, for electrons in PbTe, $\gamma=m_{\parallel}/m_{\perp}\approx10$, where $m_{\parallel}$ is the mass at $L$ point along {[}111{]}-type directions and $m_{\perp}$ is a smaller mass for all perpendicular orientations of quasi-momentum \cite{Ravich1970}. Therefore, Drude model for isotropic bands is no longer applicable. Unfortunately, general expressions for conductivity tensor components of PbTe-like materials, valid at arbitrary magnetic field $B$, are not known. Analytical formulas for $\sigma_{ij}(B)$ have been developed at low-field limit only $(\mu B\ll1)$, see \cite{Askerov1994}.

Nevertheless, such low-field formulas can be used to estimate the expected classical component of magnetoresistance data. In particular, for PbTe layer when $B$ is parallel to {[}001{]} direction, the increase of longitudinal resistance can be expressed as
\begin{equation}
	\varrho_{xx}(B)=\varrho_{0}(1+\delta^{2}\mu_{\perp}^{2}B^{2}),\label{eq:rhoxx}
\end{equation}
where $\varrho_{0}$ is a zero-field value, $\mu_{\perp}$ is a mobility of carriers with mass $m_{\perp}$ and $\delta=0.271$ for $\gamma=10$ %
\footnote{Assuming $B\parallel [001]$ and $\mu B \ll 1$ we obtained $\delta^2 =(\gamma^3-3\gamma+2)/[3\gamma(4\gamma^2+4\gamma+1)]$, where $\gamma=m_{\parallel}/m_{\perp}$.}.
Therefore, contrary to the single band Drude model, a non-zero, positive classical magnetoresistance is expected for PbTe and other lead chalcogenides. Formula (\ref{eq:rhoxx}) was used to calculate $R_{xx}(B)=\varrho_{xx}(B)$ of our PbTe layer. Parameter $\mu_{\perp}$ was obtained from the slope of $R_{xy}(B)$ data at $B<2$~T. However, the results are just a rough estimate, since at least two channels are present in electron transport, as discussed before.

To identify a different conductive channels, which are responsible for electrical transport in bulk and layered materials, the so-called Mobility Spectrum Analysis (MSA) is usually performed \cite{Beck1987}. For that purpose it is assumed, that conductivity tensor components can be expressed as integrals of Drude-like terms, with $en\mu$ factor replaced by a continuous function $\mathcal{S}(\mu)\geqslant0$, called \emph{mobility spectrum}
\begin{equation}
	\sigma_{xx}(B)=\int_{-\infty}^{\infty}\frac{\mathcal{{S}}(\mu)\,}{1+\mu^{2}B{}^{2}}\mathrm{\,{d}}\mu,\label{eq:sigmaxx}
\end{equation}
\begin{equation}
	\sigma_{xy}(B)=\int_{-\infty}^{\infty}\frac{\mathcal{{S}}(\mu)\,\mu B}{1+\mu^{2}B{}^{2}}\mathrm{{d}}\mu.\label{eq:sigmaxy}
\end{equation}

This fundamental claim is based on the model of classical transport developed by McClure \cite{McClure1956}, who solved Boltzmann equation, in the presence of magnetic field and for materials with arbitrary shape of the Fermi surface. In his approach, current carriers travel in reciprocal space on a closed constant energy curve, called \emph{hodograph}, which lays on a plane perpendicular to the direction of $B$. Formulas \ref{eq:sigmaxx} and \ref{eq:sigmaxy} are valid, if scattering relaxation time $\tau$ is constant on the hodograph, which is a \emph{basic assumption} of mobility spectrum analysis. The application of MSA method lies in finding function $\mathcal{S}(\mu)$ from experimental data.

The shape of the mobility spectrum $\mathcal{S}(\mu)$ provides a deeper insight into transport mechanisms present in the conducting sample and delivers more information as compared to the resistivity data alone. Usually, separate spectral peaks are interpreted as distinct conduction channels related to electrons $(\mu<0)$ in conduction band or holes $(\mu>0)$ in valence band. An additional information about surface conductivity, impurity bands or interface transport channels in layered structures can be also inferred from MSA \cite{Antoszewski1995}.

What is less appreciated, mobility spectrum may reveal the electron-like and hole-like peaks also for the \emph{single-band} transport, when Fermi surface is \emph{warped}. As explained in \cite{McClure1956}, convex and concave parts of the hodograph deliver distinct contributions to $\mathcal{S}(\mu)$, which differ by the sign of the related mobility. For example, not only a strong hole-like peak but also a weak feature for $\mu<0$, are expected in the mobility spectrum of \emph{p}-type silicon \cite{Dresselhaus1955}. It is clear from Fig.~\ref{fig:snte-dirac-cones} that even stronger effects, related to the shape of Fermi surface, are expected for SnTe topological states which exist on the (001) plane.

\subsection{McClure model for topological states\label{sec:McClure-model-for}}

We have applied McClure model of classical transport \cite{McClure1956} to 2D carriers with dispersion relation $E(k_{x},k_{y})$ shown in Fig.~\ref{fig:snte-dirac-cones}. The model introduces the cyclotron frequency $\omega_{c}$, defined as $2\piup/T_0$, where $T_0$ is the period of oscillatory motion performed on the hodograph. Period and frequency were calculated numerically for given energy $E$ and magnetic field $B$ by solving coupled differential equations for $k_x(t)$ and $k_y(t)$, where $t$ is time. From the solution we calculated group velocities $v_x(t)$ and $v_y(t)$, by using the dispersion relation for topological carriers again. Finally, we obtained $\sigma_{xx}(\omega_c)$ and $\sigma_{xy}(\omega_c)$ by applying Fourier transforms, as explained in \cite{McClure1956}. Tensors for a single hodograph  are of course anisotropic, however, averaging over all four Dirac-like cones, restores the cubic symmetry.

To obtain conductivity tensor components as a function of magnetic field, we took $\mu\,B=\omega_{c}\tau$, where $\tau$ is the scattering relaxation time, which in general may depend on the energy $E$ and momentum $k_{z}$ (in our case of 2D carriers, $k_{z}=0$). For calculations we used $B_{\mathrm{max}}=1\ \mathrm{T}$ and since scattering time is not known, we used an arbitrary value $\tau=5 .523\ \mathrm{ps}$, just to obtain the condition $\mu B = 1$ for $B\approx 0.5\ \mathrm{T}$. Details of the calculations will be described elsewhere, here the representative results for selected values of energy $E$ and temperature $T=0$ are presented. We have used normalized values of conductivity tensor components $s_{ij}=\sigma_{ij}/\sigma_{0}$ and dimensionless magnetic field parameter $b=B/B_{\mathrm{max}}$, where $\sigma_{0}=\sigma_{xx}(0)$ and $B_{\mathrm{max}}$ is the maximum value of $B$. By common convention we assumed $\mu<0$, $\sigma_{xy}<0$ for electrons and $\mu>0$, $\sigma_{xy}>0$ for holes.

\begin{figure*}
	\begin{centering}
		\includegraphics{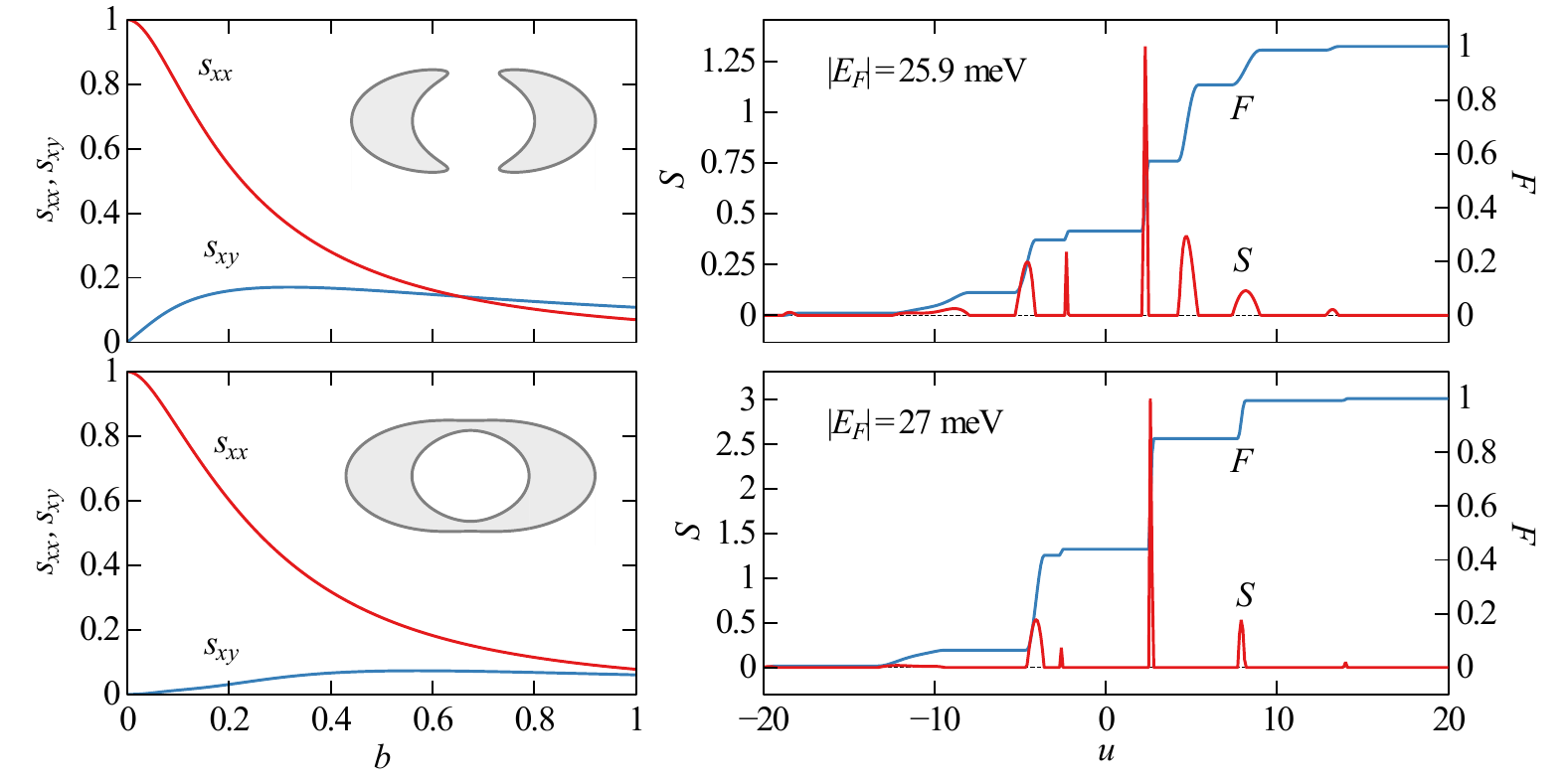}
		\par\end{centering}
	\caption{Normalized conductivity tensor components (left) for Fermi energies
		$25.9\ \mathrm{meV}$ (upper row) and $27\ \mathrm{meV}$ (below),
		insets show schematically shapes of hodographs. Corresponding mobility
		spectra $S$ and cumulative conductivity distribution functions $F$
		are show on the right. Parameter $b=B/B_{\mathrm{max}}$ is the normalized
		magnetic field, $u=\mu\,B_{\mathrm{max}}$ is the normalized mobility.
		For calculations we used $\tau=5.523\ \mathrm{ps}$ and $B_{\mathrm{max}}=1\ \mathrm{T}.$\label{fig:msa_25_27}}
\end{figure*}

Figure~\ref{fig:msa_25_27} shows $s_{xx}$ and $s_{xy}$ calculated for $|E_{\mathrm{F}}|=25.9\ \mathrm{meV}$ and $|E_{\mathrm{F}}|=27\ \mathrm{meV}$. The first case corresponds to the situation when sample is slightly below Lifshitz transition (see Fig.~\ref{fig:snte-dirac-cones}a), the second case corresponds to energies which are above transition point (Fig.~\ref{fig:snte-dirac-cones}b). Insets show closed curves in reciprocal space (hodographs), on which current carriers are orbiting in magnetic field. Clearly, the obtained data cannot be described by the Drude model of a spherical band. In both cases $\sigma_{xy}$ components of conductivity tensor are rather small, because the convex and concave parts of hodographs give contributions of opposite sign, which partially cancel each other. This mutual cancellation is stronger for energies $E>\delta$, when a separate hodographs for ``holes'' and ``electrons'' exist.

Specially interesting is the situation for $|E_{\mathrm{F}}|$ slightly
below $\delta=26\ \mathrm{meV}$, when all hodographs have a crescent-like shape. In that case ``hole'' transforms to ``electron'' and then becomes ``hole'' again, as a function of time. Moreover, in the real space a current carrier orbits in clockwise and then anti-clockwise direction or vice versa. It means that quasi-particle performs a loop, which crosses at a single point on $xy$ plane, provided $\mu B>1$ %
\footnote{The motion on a hodograph in momentum space can by converted to an orbit in real space by calculating group velocities $v_x=\partial E(k_x, k_y)/\partial k_x$, $v_y=\partial E(k_x, k_y)/\partial k_y$ and plotting the parametric curve $x(t)=v_x(t)\,t$, $y(t)=v_y(t)\,t$, where $t$ is time.}. 
Moreover, such a complicated band carrier dynamics cannot be described by simply adding two Drude-like terms, one for ``electrons'' and one for ``holes''. It was confirmed by the mobility spectrum analysis of calculated conductivity tensors.

\subsection{MSA for topological states}

To obtain mobility spectrum one has to solve the integral equations \ref{eq:sigmaxx} and \ref{eq:sigmaxy}. This is not an easy task since $\mathcal{S}(\mu)$ often consists of a few narrow peaks, whereas $\sigma_{xx}$ and $\sigma_{xy}$ are rather smooth functions of magnetic field $B$. Therefore, several numerical approaches to MSA exist in the literature \cite{Beck1987,Antoszewski1995,Chrastina2003}. Here, by replacing integrals \ref{eq:sigmaxx} and \ref{eq:sigmaxy} by discrete sums (the trapezoidal rule), we allowed iterative procedures to perform least-square fits. For the latter task, we adopted constrained optimization algorithms available in the SciPy module of the Python ecosystem \cite{Virtanen2020}, requiring that spectrum function $\mathcal{S}(\mu)$ is non-negative. Furthermore, we adopted a special mathematical procedures to avoid spurious splitting of spectral lines and the obtained results were cross-checked by independent calculations with additional constraints put on the number of allowed peaks. The details of our numerical approach to MSA are prepared for publication.

We applied our method to the normalized values of conductivity tensor components $s_{ij}=\sigma_{ij}/\sigma_{0}$, calculated above. We obtained normalized spectrum $S=\mathcal{S}\:/(\sigma_{0}B_{\mathrm{max}})$ as a function of dimensionless mobility parameter $u=\mu\:B_{\mathrm{max}}$. From \ref{eq:sigmaxx} it is clear that for $b=0$
\begin{equation}
	\int_{-\infty}^{\infty}S(u)\,{d}u=1.
\end{equation}
Therefore, in analogy to probability theory, we defined \emph{cumulative conductivity} distribution
\begin{equation}
	F(u)=\int_{-\infty}^{u}S(\xi)\,{d}\xi,
\end{equation}
which typically shows step-like behavior, each step height gives a contribution of given carrier ``species'' to the total conductivity. Results of MSA calculations for topological states are shown in Fig.~\ref{fig:msa_25_27}.

Mobility spectrum $S$ obtained for holes with $|E_{\mathrm{F}}|=25.9\ \mathrm{meV}$ contains two narrow peaks which are quite symmetric around $u=0$ and are clearly identified as the contributions arising from convex and concave parts of the sickle-shaped hodograph. However, their shares in the total conductance are not dominant, as can be seen from $F(u)$ distribution function. For example, the strongest hole-like peak for $u\approx2.3$ amounts for only 26 \% of the total conductivity. Interestingly, a dominant contribution to electrical transport comes from an additional series of wider maxima, which are observed for higher positive and negative mobilities.

\begin{figure*}
	\begin{centering}
		\includegraphics{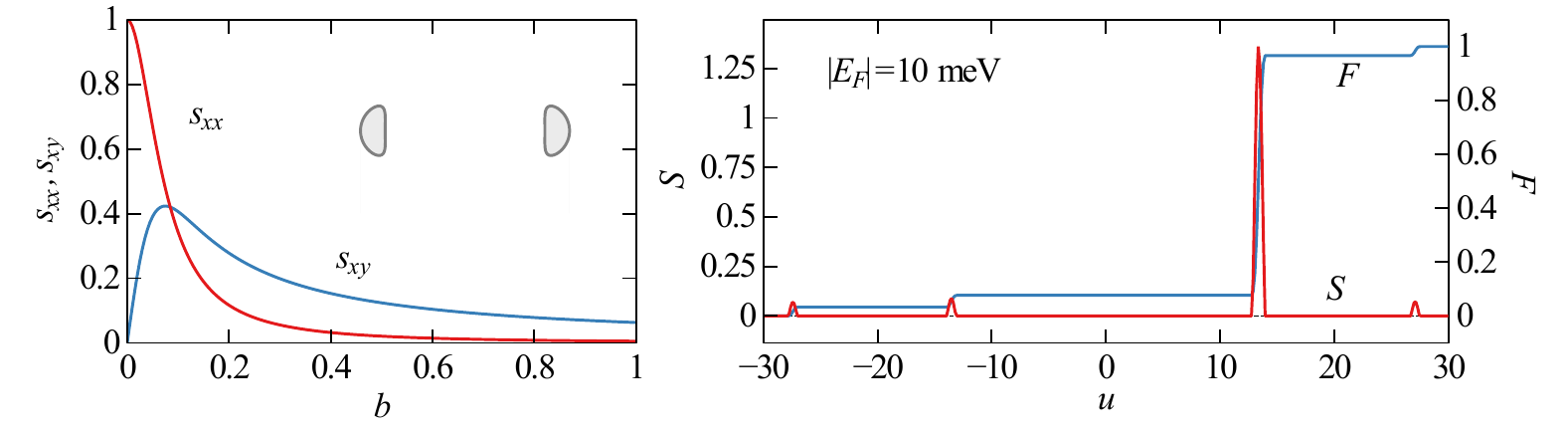}
		\par\end{centering}
	\caption{Normalized conductivity tensor components (left) for Fermi energy
		$10.0\ \mathrm{meV}$, inset shows schematically shape of hodographs.
		Mobility spectra $S$ and cumulative conductivity distribution functions
		$F$ are show on the right, we used $\tau=5.523\ \mathrm{ps}$ and
		$B_{\mathrm{max}}=1\ \mathrm{T}$, the same as in Fig. \ref{fig:msa_25_27}.\label{fig:msa_10}}
\end{figure*}

Those \emph{satellite peaks} are related to the higher harmonics of periodic cyclotron motion performed on a strongly warped orbit. Therefore, such contribution decreases when warping is smaller. Figure \ref{fig:msa_25_27} shows also the mobility spectrum obtained for $|E_{\mathrm{F}}|=27\ \mathrm{meV}$. In this case, we have two separate orbits for carriers moving in opposite directions. Nevertheless, Drude-like model for electron- and hole-like species is again not adequate since strong additional satellite peaks are still detected by MSA. However, the number of ``spectral lines" and their positions change considerably, when Fermi level  crosses Van Hove singularity.

What is more, the contribution of higher harmonics does not vanish completely, even when Fermi energy $E_{\mathrm{F}}$ approaches Dirac point. Figure \ref{fig:msa_10} shows mobility spectrum $S$ obtained for holes with $|E_{\mathrm{F}}|=10\ \mathrm{meV}$. As expected, a strong \emph{p}-type maximum dominates the conductivity, however, an \emph{n}-type component is also present together with weaker satellite peaks, visible on both electron and hole sides. As indicated in figure caption, spectrum was calculated with the same relaxation time as before. Nevertheless, carrier mobilities are much higher as compared to the case when Fermi energy is larger. This is because of \emph{linear energy dispersion} close to the Dirac point, where cyclotron frequency $\omega_{c}$ diverges as $1/E_{\mathrm{F}}$ . Such dramatic behavior should be taken into account in MSA of topological materials, because it is in stark contrast with normal matter, where $\omega_{c}$ does not depend on the energy at the band edge.

\subsection{Mobility spectrum of SnTe/PbTe junctions\label{sec:Mobility-spectrum-of-junctions}}

We applied our method of mobility spectrum calculation to analyze experimental data described in Sec.~\ref{results}, in order to identify contributions from topological states. Since PbTe and SnTe crystallize in cubic crystal structures with very similar lattice constants, interfaces are only slightly stretched. Therefore we assumed that such biaxial strain will not introduce a noticeable anisotropy to transport data, which may influence the results of MSA if $\sigma_{xx} \neq \sigma_{yy}$ \cite{Stephenson2018}. We did not, however, expect a high resolution spectra, similar to that visible in Fig.~\ref{fig:msa_25_27}, because of two reasons. Firstly, the total conductivity is dominated by bulk (3D) electrons and holes, residing outside \emph{p--n} junction. Secondly, we observed $\sigma_{xx}\gg\sigma_{xy}$ for all available data up to $B_{\mathrm{max}}=15\ \mathrm{T}$. This indicates, that a dominant contributions to electronic transport come from rather low-mobility ($\mu B_{\mathrm{max}}\ll1$) carrier species, which reduces the resolution of MSA method.

\subsubsection{MSA of PbTe layer}

\begin{figure}
	\begin{centering}
		\includegraphics{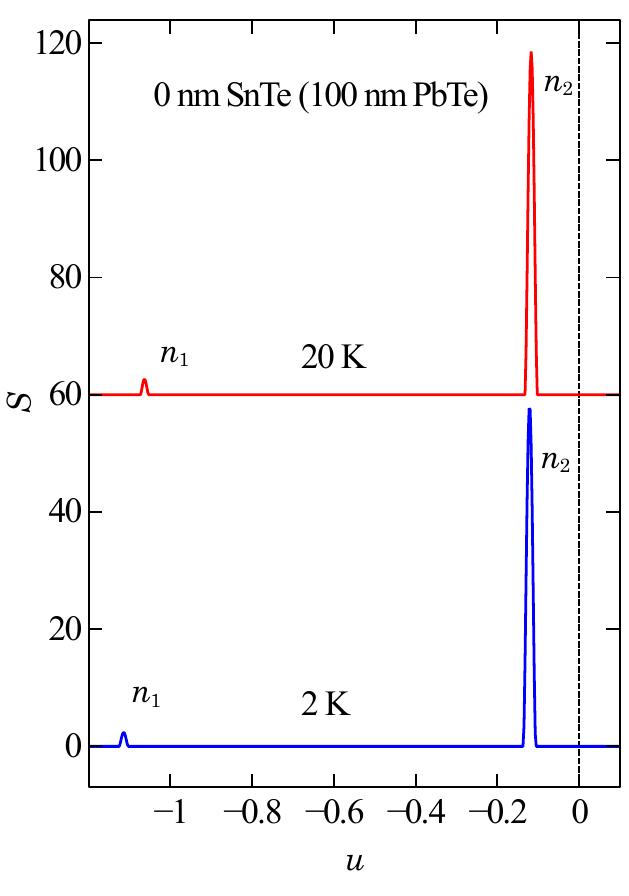}
		\par\end{centering}
	\caption{Normalized mobility spectra $S=\mathcal{S}\:/(\sigma_{0}B_{\mathrm{max}})$
		for PbTe layer at temperatures $2\ \mathrm{K}$ and $20\ \mathrm{K}$
		as a function of dimensionless mobility parameter $u=\mu\:B_{\mathrm{max}}$
		($u=1.0$ corresponds to $\mu=666.7\ \mathrm{cm^{2}}/\mathrm{Vs}$,
		since $B_{\mathrm{max}}=15\ \mathrm{T}$). Data for $T=20\ \mathrm{K}$
		are shifted up for clarity. Two electron-like peaks are indicated
		as $n_{1}$ and $n_{2}$ \label{fig:spec-PbTe}}
\end{figure}

The additional difficulty, related to PbTe data, arises because of weak localization effects, which mask the expected positive magnetoresistance in the whole range of magnetic fields. Therefore, to perform MSA, we assumed that $R_{xx}(B)=\rho_{xx}(B)$ is given by the formula \ref{eq:rhoxx} with parameters estimated from the $R_{xy}(B)$ curve. From that estimate we calculated conductivity tensor components and then mobility spectra. In order to increase the weight of $\rho_{xy}(B)$ data, which are not estimated but taken directly from experiment, we fitted the numerical derivative $\partial\sigma_{xy}/\partial B$ instead of $\sigma_{xy}$. Results are shown in Fig.~\ref{fig:spec-PbTe}.

Clearly, two electron-like peaks $n_{1}$ and $n_{2}$ are visible at temperature $T=2\ \mathrm{K}$. The dominant one, observed at mobility $\mu_{2}=-80.9\ \mathrm{cm^{2}}/\mathrm{Vs}$, is obviously related to PbTe bulk electrons with sheet carrier concentration estimated to $n_{2}=\num{3.26e13}\ \mathrm{cm^{-2}}$. The weaker peak, detected for much higher electron mobility $\mu_{1}=-743\ \mathrm{cm^{2}}/\mathrm{Vs}$, is apparently related to the 2D electron gas (2DEG) with density $n_{1}=\num{1.07e11}\ \mathrm{cm^{-2}}$, which is responsible for WL and WAL effects, as discussed before. Most probably, 2DEG resides at the free PbTe surface, as no quantum transport was reported for PbTe/CdTe interfaces grown on (100) substrates \cite{Karczewski2015}. Using well-known expressions for conductivity and density-of-states effective masses \cite{Ravich1970}, we have estimated elastic mean free paths for bulk ($n_{2})$ and surface $(n_{1})$ electrons as $\ell_{2}=1.02\ \mathrm{nm}$ and $\ell_{1}=1.33\ \mathrm{nm}$ respectively. Both values are similar and rather small, which suggest the presence of a structural disorder, which significantly reduces the mobility of carriers.

Figure \ref{fig:spec-PbTe} shows also the mobility spectrum for $T=20\ \mathrm{K},$ which looks very similar to the results obtained at lower temperature. The widths and heights of observed peaks are almost unchanged, which indicates that both carrier densities only weakly change with the temperature. As regards mobilities, $\mu_{2}$ for bulk electrons is also almost unchanged, whereas mobility of 2DEG is slightly reduced to $\mu_{1}=-709\ \mathrm{cm^{2}}/\mathrm{Vs}$. Nevertheless, the temperature dependence of layer conductivity components is rather weak in the range $2\ \mathrm{K}$ to $20\ \mathrm{K}$.

The physical origin of the two-dimensional conduction, described by parameters $n_{1}$ and $\mu_{1}$, is not known. Most probably, the presence of 2DEG is related not only to classical surface states but also to states associated with molecules chemically absorbed from air. In particular it is known, that oxygen molecules draw electrons from the bulk of semiconductor via oxidation processes. Therefore, a thin insulating layer may be formed and act as a confining barrier for electrons on the free surface of $n-$PbTe \cite{Aleksandrova2000}.

\subsubsection{MSA for SnTe/PbTe junctions}

For SnTe/PbTe junctions we did not observe weak localization, which dominated $R_{xx}(B)$ data for PbTe layer. Instead, we observed a positive magnetoresistance and the temperature dependent narrow minima at low magnetic fields, related to WAL. Therefore, to calculate components of classical conductivity tensor, we replaced low-field $R_{xx}$ data (dominated by quantum effects) with the parabolic fit. Magnetoresistance data for higher fields, i.e. for $|B|\gtrsim2\ \mathrm{T}$, were not fitted and were taken directly from experiment. As regards Hall resistance $R_{xy}(B)$, data were not fitted nor modified, even at low-field range, since the corrections related to quantum interference does not contribute to the Hall effect \cite{Fukuyama1980a,Altshuler1980} and the quantum effects related to electron-electron interactions (EEI) are negligible in our case. Indeed, for EEI effect, the normalized correction to the Hall coefficient is proportional to the relative change of sheet resistance $\Deltaup R_{xy}/R_{xy} = \xi\ \Deltaup R_{xx}/R_{xx}$, where $0 < \xi < 2$ indicates the coexistence of EEI and W(A)L effects \cite{Altshuler1980a}. By direct comparison we observed, that for our data $\xi \rightarrow 0$ which means that quantum corrections are dominated by interference phenomena.

\begin{figure}
	\begin{centering}
		\includegraphics{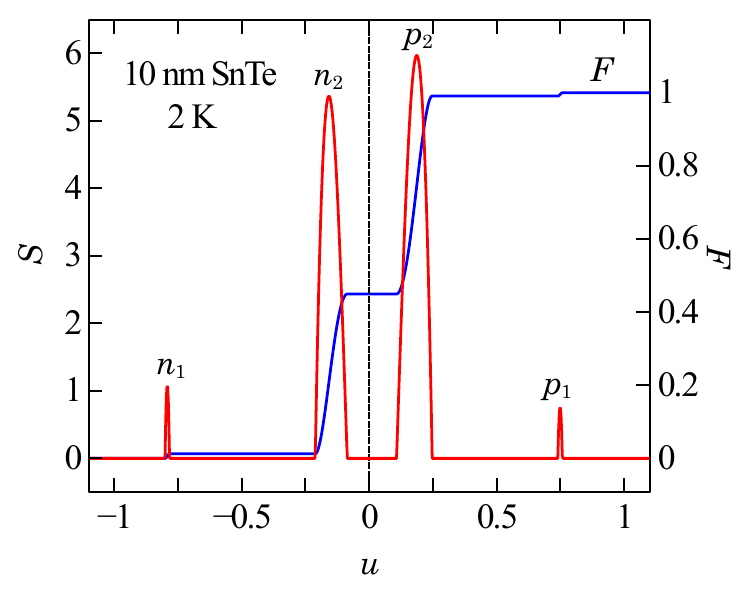}
		\par\end{centering}
	\caption{Normalized mobility spectrum $S$ and step-like cumulative conductivity
		distribution functions $F$ for 10 nm SnTe/ PbTe junction, temperature
		$T=2\ \mathrm{K}$. From steps height the following averaged parameters
		were estimated: \emph{electron-like} $\mu_{1}=-528\ \mathrm{cm^{2}}/\mathrm{Vs}$,
		$n_{1}=\num{0.85E11}\ \mathrm{cm^{-2}}$, $\mu_{2}=-105\ \mathrm{cm^{2}}/\mathrm{Vs}$,
		$n_{2}=\num{1.45E13}\ \mathrm{cm^{-2}}$ and \emph{hole-like} $\mu_{1}=+500\ \mathrm{cm^{2}}/\mathrm{Vs}$,
		$p_{1}=\num{0.59E11}\ \mathrm{cm^{-2}}$, $\mu_{2}=+125\ \mathrm{cm^{2}}/\mathrm{Vs}$,
		$p_{2}=\num{1.47E13}\ \mathrm{cm^{-2}}$. \label{fig:spec-10nm-2K}}
\end{figure}

The $\sigma_{xx}(B)$ and $\sigma_{xy}(B)$ experimental curves  calculated in this way were then used to obtain the mobility spectra. Results for $10\ \mathrm{nm}$ SnTe junction at temperature $T=2\ \mathrm{K}$ are shown in Fig.~\ref{fig:spec-10nm-2K}.
As expected, two dominant contributions are observed, which corresponds to bulk electrons in PbTe ($n_{2})$ and bulk holes in SnTe ($p_{2})$ layers. However, the widths of both spectral peaks are much larger as compared to $n_{2}$ line shown in Fig.\ \ref{fig:spec-PbTe} for single PbTe layer. We believe, that the dominant spectral lines are wider for SnTe/PbTe samples due to the formation of \emph{p--n }heterojunction at the interface. Electron and hole densities vary across junction and the same may apply to mobilities, leading to the spread of parameters and spectral broadening. Interestingly, the total sheet densities $n_{2}$ and $p_{2}$ are quite similar, see Fig.\ \ref{fig:spec-10nm-2K}. However, if the thicknesses of PbTe ($100\ \mathrm{nm}$) and SnTe ($10\ \mathrm{nm}$) layers are taken into account, one obtains $n_{2}^{\mathrm{3D}}=\num{1.45E18}\ \mathrm{cm^{-3}}$ and $p_{2}^{\mathrm{3D}}=\num{1.47E19}\ \mathrm{cm^{-3}}$ for bulk carrier concentrations, which differs considerably.

\begin{figure*}
	\begin{centering}
		\includegraphics{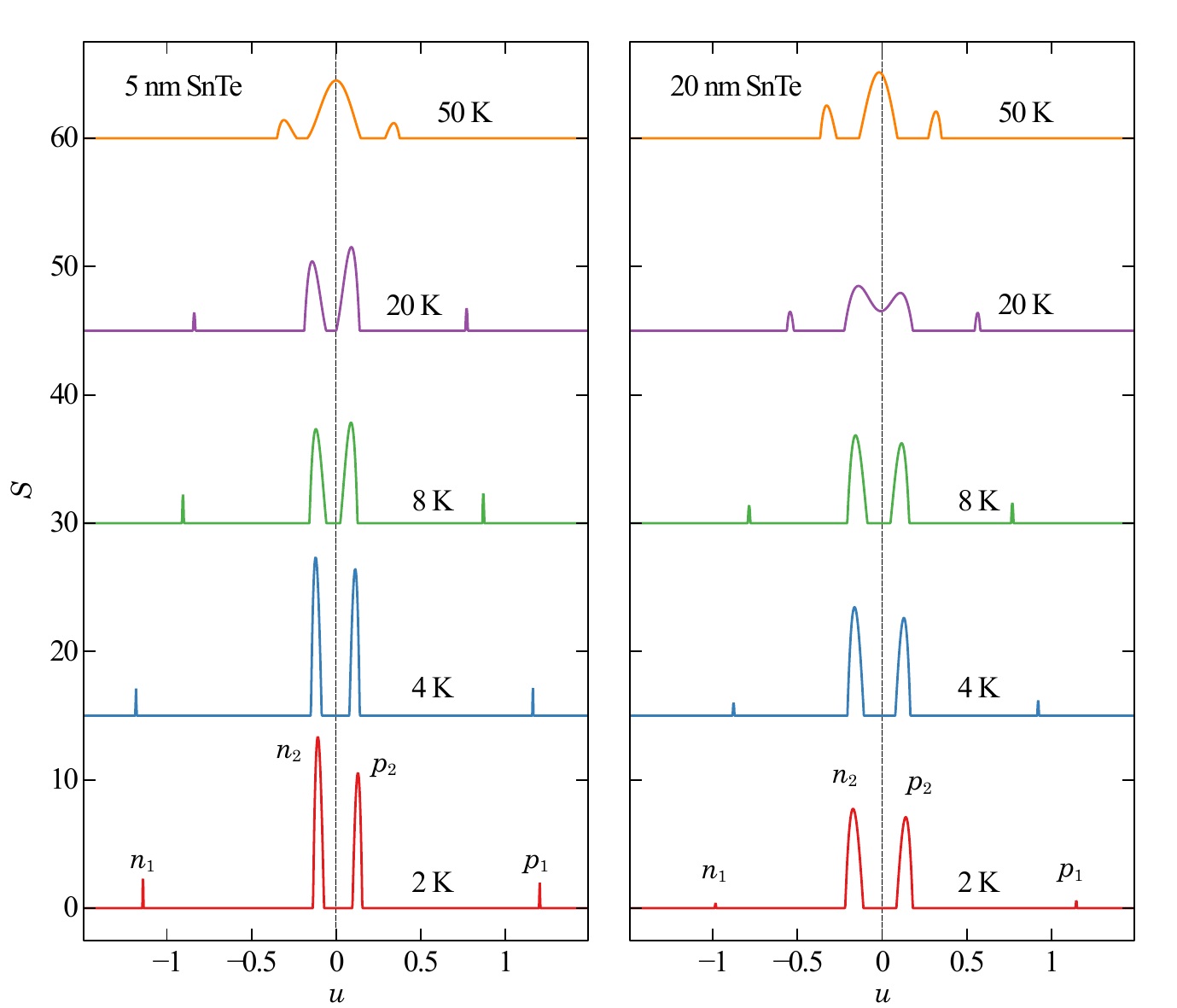}
		\par\end{centering}
	\caption{Mobility spectra for 5 nm SnTe/ PbTe and 20 nm SnTe/ PbTe junction as a function of temperature in the $2\ \mathrm{K}$ to $50\ \mathrm{K}$ range. \label{fig:spec=00003D5nm-20nm-vs-T}}	
\end{figure*}

MSA shows that the bulk densities are not the only contributions to conductivity for SnTe/PbTe samples. The additional $n-$type ($n_{1}$) and $p-$type ($p_{1})$ peaks are clearly observed. We attribute \emph{them both} to the topological states, which live at SnTe/normal matter interfaces and which are responsible for the weak anti-localization effect. Firstly, $n_{1}$ contribution cannot be connected with a free surface of PbTe, which is not exposed to air but buried in the \emph{p--n }junction. Secondly, the formation of a 2D hole gas on \emph{p--}SnTe, similar to 2DEG observed on a surface \emph{n--}PbTe, is not expected. The SnTe surface is not capped and probably also oxidized, however, the density of holes is so high ($\sim 10^{19}\ \mathrm{cm^{-3}}$) that the formation of an insulating layer is not expected. Without such a barrier, the existence of surface accumulation (or inversion) layer is not likely here because of high dielectric constant ($\epsilon=1200$). The strong screening of surface charges limits band bending to a few meV \cite{Assaf2014}, which is much less then the estimated Fermi energy $E_{\mathrm{F}}$.

Therefore, the two satellite peaks $n_{1}$ and $p_{1}$ are not related to the conventional 2D transport in trivial semiconductor. As discussed in section \ref{sec:McClure-model-for}, the shapes of constant energy contours on the (001) surface result in electron-like and hole-like contributions to mobility spectrum of SnTe topological states. Moreover, this observation is valid in relatively wide range of Fermi energies, from $|E_{\mathrm{F}}|\approx10\ \mathrm{meV}$ up to $|E_{\mathrm{F}}|\approx 75\ \mathrm{meV}$, when inner and outer Dirac cones with the same chirality develop. Our estimations show, that at least for \emph{p--n} junction area, the energy of SnTe surface carriers falls within that range. Therefore, we interpret the simultaneously observed peaks $n_{1}$ and $p_{1}$ as the signatures of non-trivial chirality of topological states. This conclusion is supported by the temperature dependence of mobility spectra, which was studied for $5\ \mathrm{nm}$ SnTe and $20\ \mathrm{nm}$ SnTe heterojunctions in the temperature range $2\ \mathrm{K}$ to $50\ \mathrm{K}$. Results are shown in Fig.\ \ref{fig:spec=00003D5nm-20nm-vs-T}.

\begin{figure*}
	\begin{centering}
		\includegraphics{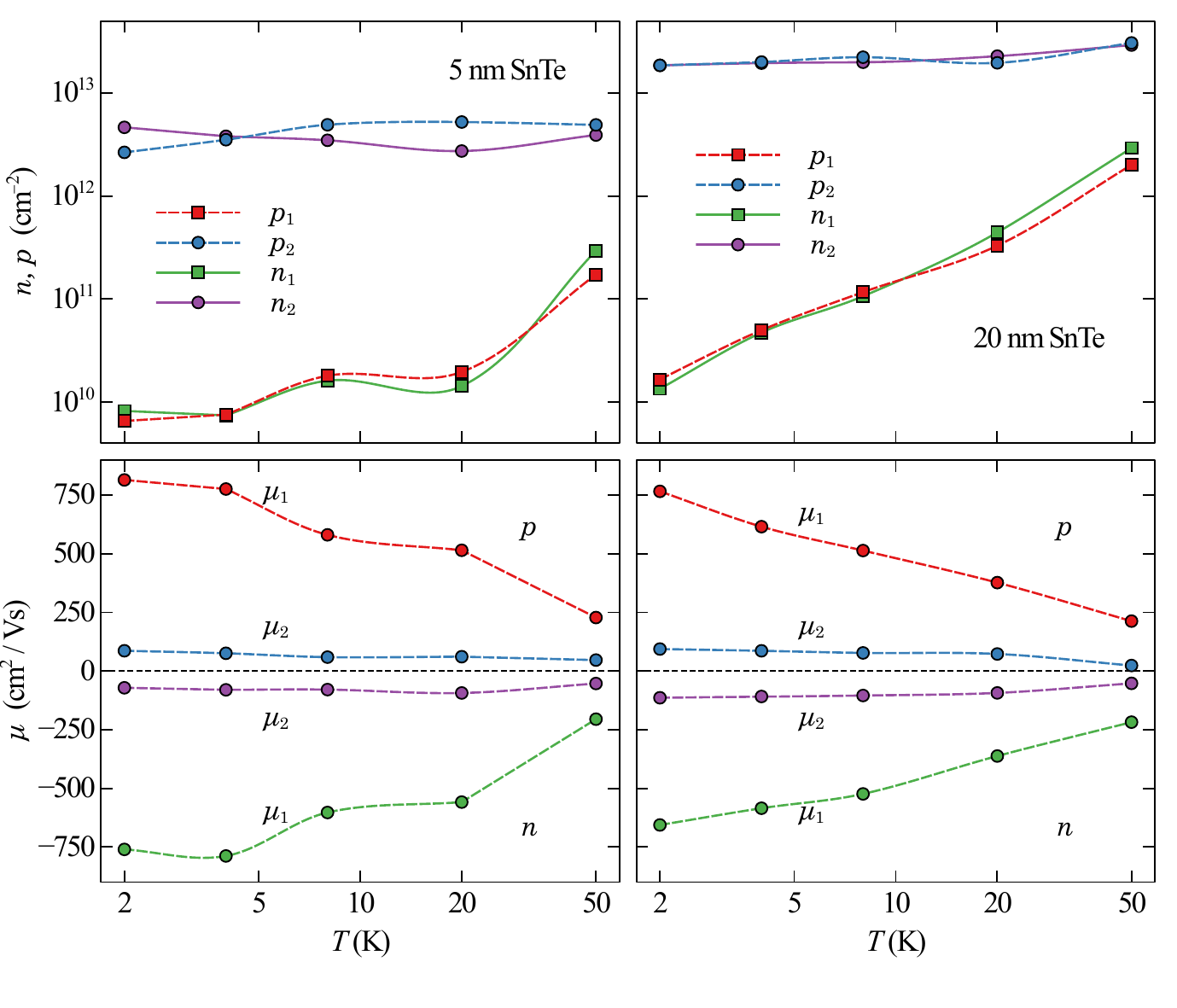}
		\par\end{centering}
	\caption{Results of MSA analysis performed for 5 nm SnTe/PbTe (left) and 20
		nm SnTe/PbTe (right) junctions. Upper part shows 2D densities of electron-like
		($n_{1}$, $n_{2}$) and hole-like ($p_{1}$, $p_{2}$) contributions
		as a function of temperature $T$. Below, the corresponding mobilities
		(peaks positions) are shown. By convention, $\mu$ for electron-like
		excitations are negative. \label{fig:MSA-5nm-20nm}}
\end{figure*}

A much stronger temperature dependence is observed as compared to PbTe layer data, which did not change much from $2\ \mathrm{K}$ to $20\ \mathrm{K}$, see Fig\ \ref{fig:spec-PbTe}. For PbTe, electron density $n_{1}$ of 2DEG increased by $7$\ \% and the absolute value of mobility $\mu_{1}$ decreased by $5$\ \% only. In contrast, for SnTe/PbTe junctions, the already large widths of peaks $n_{2}$ and $p_{2}$, which are related to bulk carriers, increased further with temperature. At $T=20\ \mathrm{K}$ the dominant spectral lines started to overlap and eventually merge at $T=50\ \mathrm{K}$, forming a broad peak, which spans over positive and negative mobilities. The satellite spectral lines $n_{1}$ and $p_{1}$ also changed considerably with $T$. For 20\ nm SnTe sample, electron-like density $n_{1}$ increased $33$ times and the absolute value of mobility $\mu_{1}$ decreased by almost $50$\ \%, when heterojunction was warmed from $2\ \mathrm{K}$ to $20\ \mathrm{K}$. This is in stark difference with 2DEG peak $n_{1},$ observed for PbTe layer. The temperature dependence of MSA parameters, obtained for SnTe/PbTe junctions, is summarized in Fig.~\ref{fig:MSA-5nm-20nm}.

The upper part of the figure shows estimated sheet densities of electron-like and hole-like contributions. As expected, the bulk carrier concentrations, marked with circles, did not change much with temperature. For 20~nm sample, parameters $n_{2}$ and $p_{2}$ slightly increased with temperature, whereas for 5~nm sample, changes were larger, but not monotonic. The same observations apply for positive and negative mobilities of majority carriers $\mu_{2}$, shown in the lower part of Fig.~\ref{fig:MSA-5nm-20nm}. As already mentioned, the widths of mobility peaks increased with temperature, but their positions also did not change considerably. Only at $T=50\ \mathrm{K}$, the absolute values of $\mu_{2}$ markedly decreased and peaks have merged. In spite of broadened spectral lines, the majority carriers in SnTe/PbTe junctions are characterized by relatively small values of mobilities with weak temperature dependence, very much as bulk electrons in PbTe layer.

However, the estimated bulk densities of electrons in SnTe/PbTe heterojunctions are considerably smaller then in the single PbTe layer. At temperature $T=2\ \mathrm{K}$ we obtained $n_{2}^{\mathrm{3D}}=\num{1.45E18}\ \mathrm{cm^{-3}}$ and $n_{2}^{\mathrm{3D}}=\num{1.87E18}\ \mathrm{cm^{-3}}$ for 10~nm and 20~nm samples respectively, as compared to $n_{2}^{\mathrm{3D}}=\num{3.26E18}\ \mathrm{cm^{-3}}$ in PbTe layer. For 5\ nm SnTe/PbTe sample, which has the largest zero-field resistance, this discrepancy is even more dramatic, since $n_{2}^{\mathrm{3D}}=\num{0.46E18}\ \mathrm{cm^{-3}}$. One possible explanation is that part of each PbTe layer was depleted due to the formation of \emph{p-n} junction and the effective widths were smaller than nominal 100\ nm value, used in calculations. Secondly, McClure model, which is the foundation of MSA method, was developed for a possibly multi-carrier but homogeneous samples. Here, we studied the layered structures with planar heterojunction and surface conduction, with unknown vertical distribution of electrical current. Therefore, the calculated carrier concentrations may differ from an actual ones.

Nevertheless, the quantitative parameters, obtained for electron-like and hole-like surface conduction, are consistent for all SnTe/PbTe samples. Figure~\ref{fig:MSA-5nm-20nm} shows, that parameters $p_{1}$ and $n_{1}$ increase with temperature and both values change in-parallel, which strongly suggests that they belong to the same entity. If they were just the hole and electron concentration of distinct 2DEG systems, such similarity is less probable since our multi-layer structures are not symmetric along the growth direction. The same applies to $\mu_{1}$ parameters for electron-like and hole-like peaks. Their absolute values decrease with temperature with almost exactly the same rates. Mobility and density of electrons on the PbTe surface behave differently. This confirms that most probably, peaks $p_{1}$ and $n_{1}$ belong to the concave and convex parts of the single hodographs, of a type shown in Fig.\ \ref{fig:msa_25_27} for topological states. The increase of $p_{1}$ and $n_{1}$ parameters with temperature can be explained if we assume, that at $T=2\ \mathrm{K}$ Fermi level $E_\mathrm{F}$ is located slightly below the  Lifshitz transition energy $E_\mathrm{S}$. Therefore, at higher temperatures, $E_\mathrm{F}$ moves towards a van Hove singularity in the density of states and the conductance related to topological carriers increases.  

If it is true, one may ask a more general question, weather the standard MSA approach is applicable to TSS. Maybe, instead of a Drude expression $\sigma=en\mu$, a more general formula $\sigma(E)=e^{2}\mathcal{D}D(E)$ should be used, where $\mathcal{D}$ is the diffusion coefficient and $D(E)$ is the density of states. Such expression is far more generally applicable and provides better descriptions for materials like topological insulators, where the meaning of effective mass and mobility is somehow unclear \cite{Datta2012}.
	
\section{Conclusions and summary}

The SnTe/PbTe junction at low temperatures is a so-called type-II heterostructure, with the valence band edge of SnTe located higher than the bottom of conduction band in PbTe \cite{Takaoka1986}. Owning to the large hole density it is expected, that electrons diffuse across the interface, SnTe bands bend downward and chemical potential is shifted towards the energy gap \cite{Wei2018}. Moreover, due to the lattice mismatch, SnTe/PbTe heterogunction is strained  and the resulting in-plain strain is compressive for PbTe and tensile for SnTe layers. As a result, for (001) junctions, the positive band offset is reduced. For sufficiently large strain, band gap is not broken anymore \cite{Litvinov1993}, which in turn promotes the formation of midgap gapless states with linear dispersion \cite{Volkov1985}.

Bi-axial strain influences also the 2D band structure of topological carriers. The tensile distortion of thin (100) SnTe film shifts the Dirac points in reciprocal space and eventually opens the hybridization gap \cite{Qian2015}. Such a situation may have arisen for 5~nm SnTe/Pbte sample, since in this case, the quantum corrections to conductance are better described by formula given by Eq.~\ref{eq:HLN}, which is valid for gaped states. Another possibility, which explains the coexistence of WAL and WL effects for this sample, is the formation of a trivial Volkov--Pankratov massive states, which are predicted in wide enough junctions due to the presence of electric field \cite{Volkov1985, Mahler2021}.

The phase coherence effects, observed for 10~nm and 20~nm SnTe/PbTe samples were fully described by a simpler Eq.~\ref{eq:HLN-Dirac}, which is valid for gapless states. Probably, the tensile strain was more relaxed for thicker SnTe layers, hybridization did not occur and gap remained broken. The effective number of quantum channels, $\eta$, was reduced for all devices, presumably by the strong inter-valley coherent scattering. For all samples, the parameter $\eta$ was further reduced due to the contribution from weak localization effect, which is expected for bulk but mesoscopic SnTe and PbTe layers.

For SnTe/PbTe junctions, the mobility spectrum analysis (MSA) revealed the presence of bulk electrons from PbTe and bulk holes from SnTe layers, as expected. Additionally, the weak hole-like and electron-like peaks  were observed for higher mobilities, always in pairs, on symmetrical positions around $\mu=0$ point. The distance between those peaks decreased with temperature, whereas their density parameter increased. The additional peak pairs were interpreted as the mobility spectrum of carriers occupying the \textit{single topological band}, with Fermi level pinned in the vicinity of Lifshitz transition ($E_\mathrm{S}=26\ \mathrm{meV})$, where density of states is very large. This conclusion is supported by the numerical calculations of conductivity tensor components for the energy band shown in Fig.~\ref{fig:snte-dirac-cones}, using McClure model of classical transport.

\begin{acknowledgments}	
We acknowledge Krzysztof Dybko for the useful discussions. The research in Poland was supported by the National Science Centre (NCN) through the project   2021/41/B/ST3/03651 and in Germany by  the  Deutsche Forschungsgemeinschaft (DFG) through  the program SFB 1170 “ToCoTronics”. Work was also partially supported by the Polish National Science Center grants No. P-376/M and DEC-012/07/B/ST3/03607. MSA calculations have been performed thanks to the support of Polish Ministry of Higher Education through the program "Regional Initiative of Excellence", grant No. 014/RID/2018/19.
\end{acknowledgments}

%

\end{document}